\documentclass[showpacs,preprintnumbers,amsmath,amssymb,10pt]{revtex4}
\usepackage{dcolumn}% Align table columns on decimal point
\usepackage{bm}% bold math

\newcommand{\s}{\ensuremath{\psi(t,r)}}
\newcommand{\n}{\ensuremath{\nu(t,r)}}
\newcommand{\T}{\ensuremath{\theta}}

\newcommand{\pt}{\ensuremath{p_\theta}}
\newcommand{\pr}{\ensuremath{p_r}}

\newcommand{\e}{equation$\;$} 
\newcommand{\M}{\ensuremath{{\cal M}}}

\newcommand{\prz}{\ensuremath{p_{r_0}}}

\newcommand{\ptz}{\ensuremath{p_{\theta_0}}}

\newcommand{\X}{\ensuremath{{\cal X}}}

\newcommand{\R}{\ensuremath{{\Re}}}

\newcommand{\dw}{\ensuremath{{d\Omega^2_{N-2}}}}
\newcommand{\Gab}{\ensuremath{{}^NG_{ab}}}
\newcommand{\G}{\ensuremath{{}^NG_{22}}}
\newcommand{\Gii}{\ensuremath{{}^NG_{ii}}}
\newcommand{\gab}{\ensuremath{g_{ab}}}

\newcommand{\rdot}{\ensuremath{\dot{R}}}
\newcommand{\rddot}{\ensuremath{\ddot{R}}}

\newcommand{\F}{\ensuremath{F(r)}}
\newcommand{\f}{\ensuremath{f(r)}}

\begin{document}

\preprint{}

\title{Spherical gravitational collapse in N-dimensions}

\author{Rituparno Goswami}
\email{goswami@tifr.res.in}
\author{Pankaj S Joshi}
\email{psj@tifr.res.in}
\affiliation{Tata Institute of Fundamental Research\\
Homi Bhabha Road, Colaba\\ Mumbai 400 005, India}

\begin{abstract} 
We investigate here spherically symmetric gravitational collapse 
in a spacetime with an arbitrary number of dimensions and with a general
{\it type I} matter field, which is a broad class that includes 
most of the physically reasonable matter forms. We show that given the initial
data for matter in terms of the initial density and pressure profiles at
an initial surface $t=t_i$ from which the collapse evolves, there exist
rest of the initial data functions and classes of solutions of Einstein 
equations which we construct here, such that the spacetime evolution goes
to a final state which is either a black hole or a naked singularity,
depending on the nature of initial data and evolutions chosen, and subject 
to validity of the weak energy condition.  The results are discussed and 
analyzed in the light of the cosmic censorship hypothesis in black hole 
physics. The formalism here combines the earlier results on gravitational 
collapse in four dimensions in a unified treatment. Also the earlier 
work is generalized to higher dimensional spacetimes to allow a study 
of the effect of number of dimensions on the possible final outcome of 
the collapse in terms of either a black hole or naked singularity. 
No restriction is adopted on the number of dimensions, and other limiting
assumptions such as self-similarity of spacetime are avoided, in order 
to keep the treatment general. Our methodology allows to consider to 
an extent the genericity and stability aspects related to the occurrence 
of naked singularities in gravitational collapse. 
\end{abstract}


\maketitle

\section{Introduction}

A considerable amount of work has continued in recent years
on the validity or otherwise of the cosmic censorship conjecture (CCC) in
black hole physics. The reason
for this interest is that CCC is fundamental to many aspects of 
the basic theory and
applications of black holes, and the astrophysical implications resulting
from the phenomena of continual gravitational collapse of a massive 
star which has exhausted its nuclear fuel. As of today, no theoretical 
proof, or even any satisfactory mathematical formulation of CCC is 
available, where as an intriguing finding that has emerged from recent 
investigations on gravitational collapse scenarios in general relativity is 
that the final end state of such a collapse could be either a 
black hole (BH), or a naked singularity (NS). 
The theoretical and observational properties
of these objects could be quite different from each other and so it is 
of much interest to get an insight into how each of these phases come 
about as end states of a dynamically developing collapse governed  
by gravitational dynamics.

When a sufficiently massive star starts 
collapsing gravitationally on exhausting its nuclear fuel, it would 
not settle to a stable configuration such as a neutron star. What 
happens in such a case is an endless gravitational collapse ensues, where 
the sole governing force is gravity. According to general relativity,
the outcome of such a process would be necessarily a spacetime 
singularity. If an event horizon of gravity forms well in advance 
before the singularity forms that gives rise to a black hole as the 
final state of collapse. In the case 
otherwise, when the horizon formation is delayed during the collapse, 
these extreme density and curvature regions may fail to be covered 
by the horizon, and a visible naked singularity may develop (see e.g. 
~\cite{rev1} -~\cite{rev7} 
for some recent reviews). 
The possible physical consequences of the later scenario have
drawn some attention recently
~\cite{rev1}.
Most of the gravitational collapse models studied so far are 
spherical, and the matter cloud collapses under reasonable physical 
conditions such as an energy condition, and regularity conditions on 
the initial data from which the collapse develops.

A note-worthy suggestion that has emerged towards a possible 
theoretical formulation of CCC is that, any naked singularities resulting 
from matter models which may also develop singularities in special 
relativity, should not be regarded as physical
~\cite{cch1} -~\cite{cch5}.
Clearly, it will require a serious effort to cast this into a 
mathematical statement and a possible proof for CCC. Also, it may not 
be easy to discard completely all the matter fields such as dust, perfect 
fluids, and matter with various other reasonable equations of state, 
which have been studied and used extensively in relativistic astrophysics 
for a long time from the perspective of understanding gravitational 
collapse processes and final states.

Another possibility that indeed appears worth exploring
is we may actually be living in a higher dimensional (HD) spacetime. 
The recent developments in string theory and other related field 
theories strongly indicate that gravity is possibly a higher dimensional 
interaction, which reduces to the general relativistic description at lower 
energies. Hence, there is a possibility that while CCC may possibly 
fail in the 
four-dimensional manifold of general relativity, it may well be restored 
in higher dimensions due to the extra physical effects arising from 
our transition itself to a higher-dimensional spacetime continuum. Such
considerations would inspire a study of gravitational collapse in higher
dimensional spacetimes. From such a perspective, many works have
reported results in recent years on spherically symmetric collapse in HD.
The recent revival of interest in this problem is
partly motivated by the Randall-Sundrum brane-world scenario
~\cite{hdim1}
and different `sectors' of the general problem have 
been studied by restricting to certain subcases such as those
in self-similar spacetimes, or spacetimes with a specific number of 
dimensions
~\cite{hdim2} -~\cite{hdim10}.

It follows that a general investigation of
gravitational collapse in a spacetime with an arbitrary number
of higher dimensions will be of considerable interest. From the 
perspective of CCC, the effects the number of dimensions may have on 
the final outcome of gravitational collapse in terms of BH/NS phases 
will be of much interest.

From such a perspective, we investigate here the issue of 
BH/NS endstate formations in a higher dimensional gravitational 
collapse in some detail, in order to bring out in a transparent manner 
the effect of dimensions on the final fate of evolution of a matter 
cloud which collapses from a given regular matter initial data. A 
spherically symmetric collapse is considered here in $N\ge3$ 
dimensions and the matter content is chosen to be of {\it Type I}, 
which obeys the weak energy condition. The collapsing matter field we 
consider here is a general and broad class, which includes most of 
the physically reasonable matter fields such as dust, perfect fluids, 
massless scalar fields and such others and restrictions of any 
special form are not imposed on the form of matter.

We consider spherically symmetric spacetimes here, however, when 
compared to some of the earlier studies of collapse in higher dimensions 
mentioned above, we deal here with the general case. That is, 
there is no further restriction imposed on the number of dimensions, the 
spacetime 
is not assumed to obey various special conditions such as self-similarity 
or assuming the existence of various Killing fields, which are somewhat 
restrictive assumptions for collapse models. The results here also 
generalize earlier results 
~\cite{gen1}-\cite{gen4},~\cite{global},  
related to gravitational collapse final states, and provide a unified 
treatment.

In order to investigate collapse final states, given 
the initial data for matter in terms of the initial density and pressure 
profiles at an initial surface $t=t_i$ from which the collapse develops, 
we construct classes of solutions to Einstein equations, such that the 
spacetime evolution goes to a final state which is either a black hole 
or naked singularity, depending on the nature of rest of the free initial 
data functions and possible evolutions, subject to validity of the weak 
energy condition. The methodology used here allows us to consider 
in some detail the genericity and stability aspects related to the 
occurrence of naked singularities in gravitational collapse.

In the next Section II, we describe the basic equations and
also the various regularity conditions for gravitational collapse. 
Section III then considers spherical collapsing clouds and their 
basic dynamics. The apparent horizon and structure of trapped surface 
formation is discussed in Section IV, which has a direct bearing on the 
nature of singularity in terms of being either visible or covered. 
The issues related to equation of state and that of validity of energy 
conditions in our consideration are discussed in Section V, and we also 
discuss briefly collapse in lower spacetime dimensions in Section IV. 
The exterior spacetime and matching conditions are given in Section 
VII and the final Section summarizes some concluding remarks.

\section{Einstein equations, regularity and energy conditions}

Let us consider a general N-dimensional spherically symmetric 
metric of the form,
\begin{equation}
ds^2=-\gab(x^0,x^1)dx^adx^b+R^2(x^0,x^1)\dw
\label{eq:genmetric}
\end{equation}
where $a,b$  run from $0$ to $1$, and
\begin{equation}
\dw=\sum_{i=1}^{N-2}\left[\prod_{j=1}^{i-1}\sin^2(\T^j)\right](d\T^i)^2
\end{equation}
is the metric on $(N-2)$ sphere with $\T^i$ being  the spherical
coordinates.
From this metric, we get the elements of Einstein tensor as
~\cite{hdim6}~\cite{hdim7},
\begin{equation}
\Gab=\frac{N-2}{2R^2}\left[\gab\{2R\Box R+(N-3)Q\}-2RR_{,ab}\right]
\end{equation}
\begin{equation}
\G=-\frac{1}{2}\left[R^2\R+(N-3)\{2R\Box R+(N-4)Q\}\right]
\end{equation}
\begin{equation}
\Gii(i>2)=\left[\prod_{k=1}^{i-2}\sin^2(\T^k)\right]\G
\end{equation}
where we have,
\begin{eqnarray}
Q=1+R^{,a}R_{,a};&\Box R=g^{ab}R_{;ab}
\end{eqnarray}
and,
\begin{equation}
\R=g^{ab}\R_{ab}
\end{equation}
Here $\R_{ab}$ is the  Ricci tensor evaluated by the two-metric \gab, 
and  $\Re$ is the scalar curvature evaluated by the same. 
Also, throughout this paper we use the ususal convention, that is, a 
comma(,) in subscript denotes 
partial differentiation while a semicolon (;) denotes covariant differentiation.

Let us now describe the spacetime geometry within the spherically 
symmetric collapsing cloud by the comoving coordinates $(t,r,\theta^i)$,
which are specified as below.
We take the matter 
field to be of {\it Type I}, which is a broad class including 
most of the physically reasonable matter forms, including dust, 
perfect fluids, massless scalar fields and such others. 
This is specified by the requirement that the energy momentum tensor 
for the matter admits one 
timelike and $(N-1)$ spacelike eigenvectors 
~\cite{haw}.
We now choose our co-ordinates 
$(t,r,\theta^i)$ to be along these eigenvectors, which makes 
the coordinate system to be {\em comoving}, that is, the co-ordinate 
system moves with the matter. We can use
the freedom of coordinate transformations of the form 
$t'=f(t,r)$ and $r'=g(t,r)$ to make the $g_{tr}$ term in metric
(\ref{eq:genmetric}) and the radial velocity of the matter
to vanish. In that case the general metric in the comoving   
coordinates $(t,r,\theta^i)$ must have three arbitrary functions
of $t$ and $r$ and this can be written in the form
~\cite{landau},
\begin{equation}
ds^2=-e^{2\n}dt^2+e^{2\s}dr^2+R^2(t,r)\dw
\label{eq:metric}
\end{equation}
In this comoving frame the energy-momentum tensor for any matter 
field which is {\it Type I} is given in a diagonal form,
\begin{equation}
T^t_t=-\rho(t,r);\;\;T^r_r=\pr(t,r);\;\;T^{\T^i}_{\T^i}=p_{\T}(t,r)
\label{eq:em}
\end{equation}
The quantities $\rho$, $p_r$ and $p_\T$ are respectively the 
energy density, and radial 
and tangential pressures, ascribed to the matter field. 
The matter cloud has a compact support with $0<r<r_b$, where $r_b$
denotes the boundary of the cloud, outside which it is to be suitably
matched through suitable junction conditions with another 
spacetime geometry.

We take the matter field 
to satisfy the {\it weak energy condition}, that is, the energy density 
measured by any local timelike observer is non-negative. This ensures 
the physical reasonability for the collapsing matter fields we are 
considering. Another energy condition frequently used is the 
{\it dominant energy condition}, which demands that for any timelike 
observer the local energy flow is non-spacelike. We note that 
these two are frequently regarded as the main and important 
energy conditions which are physically reasonable. 
~\cite{rev3}.
For these energy conditions to be satisfied, we must have 
for any timelike vector $V^i$, 
\begin{equation}
T_{ik}V^iV^k\ge0
\end{equation}
and $T^{ik}V_k$ non-spacelike. For the energy-momentum tensor 
(\ref{eq:em}), these amount respectively to the conditions,
\begin{equation}
\rho\ge0;\; \rho+p_r\ge0;\; \rho+p_{\T}\ge0
\end{equation}
\begin{equation}
|p_r|\le\rho\;\; ; |p_{\T}|\le\rho.
\end{equation}

Now with the above metric (8), the following quantities 
can be evaluated,
\begin{equation}
R^{,a}R_{,a}=-\rdot^2e^{-2\nu}+R^{'2}e^{-2\psi}
\end{equation}
\begin{equation}
\Box R=-e^{-2\nu}[\rddot+\rdot(-\dot{\nu}+\dot{\psi})]+e^{-2\psi}
[R''+R'(\nu'-\psi')]
\end{equation}
and,
\begin{equation}
R_{,ab}=\left(\rddot-\dot{\nu}\rdot-\frac{e^{2\nu}\nu'R'}{e^{2\psi}}\right)
\delta^0_a\delta^0_b
+\left(R''-\psi'R'-\frac{e^{2\psi}\dot{\psi}\rdot}{e^{2\nu}}\right)
\delta^1_a\delta^1_b + \mathcal{Q}_{ab}
\end{equation}
where we have,
\begin{equation}
\mathcal{Q}_{ab}=\left(\rdot'-\nu'\rdot-\dot{\psi}R'\right)
(\delta^0_a\delta^1_b+\delta^1_a\delta^0_b)
\end{equation}
Using these quantities the Einstein equations $G_{ik}=T_{ik}$ take
the form, (in the units $8\pi G=c=1$),
\begin{eqnarray}
\rho=\frac{(N-2)F'}{2R^{N-2}R'}; && 
p_r=-\frac{(N-2)\dot{F}}{2R^{N-2}\dot{R}}
\label{eq:ein1}
\end{eqnarray}
\begin{equation}
\nu'=\frac{(N-2)(\pt-p_r)}{\rho+p_r}\frac{R'}{R}-\frac{p_r'}{\rho+p_r}
\label{eq:ein2}
\end{equation}
\begin{equation}
-2\dot{R}'+R'\frac{\dot{G}}{G}+\dot{R}\frac{H'}{H}=0
\label{eq:ein3}
\end{equation}
\begin{equation}
G-H=1-\frac{F}{R^{N-3}}
\label{eq:ein4}
\end{equation}
where we have defined,
\begin{eqnarray}
G(t,r)=e^{-2\psi}(R')^2; && H(t,r)=e^{-2\nu}(\dot{R})^2
\end{eqnarray}

The function $F=F(t,r)$ is known as the {\it Misner Sharp mass}, 
which gives the total mass in 
a shell of comoving radius $r$, at an epoch $t$. 
The energy condition $\rho\ge0$ imply $F\ge0$ and $F'\ge0$.
Since the area radius vanishes at the center of the cloud,
from equation (\ref{eq:ein1}) it is evident that 
in order to preserve the regularity of density and pressures 
at any non-singular epoch $t$, we must have  
$F(t,0)=0$, that is the mass function should vanish at the center  
of the cloud.

As seen from equation (\ref{eq:ein1}), there is a density 
singularity in the spacetime at $R=0$, and at $R'=0$. However, 
the later ones are due to shell-crossings
~\cite{shell1}, 
which basically indicates the breakdown 
of the coordinate system we have used. These are not generally 
regarded as genuine singularities, which can be possibly removed 
from the spacetime to extend the manifold through the same
\cite{shell2}. 
Hence we shall consider here only the shell-focusing singularity 
at $R=0$, 
which is a genuine physical singularity where all matter shells 
collapse to a zero physical radius. We shall discuss this in some 
more detail in the next section.

We note that, in general, for a general matter field
with non-vanishing pressures as we consider here, there are a 
variety of dynamical time evolutions possible from the given matter 
density and pressure profiles as prescribed on an initial surface
(which we call here matter initial data), from which the collapse
evolves. In particular, even if the cloud commences gravitational
collapse at the initial surface $t=t_i$, there 
can be classes of solutions of Einstein equations where the evolution
is such that a bounce is possible at a later stage for the cloud. 
We consider here only continually collapsing class of models, because 
our interest is in the physical situation which corresponds to the 
case when the mass of the star is so high that on exhausting its 
nuclear fuel, it must undergo a continual gravitational collapse,
completing the same in a finite time. Thus the continual
collapse condition is included as a part of our work here.  
In this case, trapped surfaces develop and a spacetime singularity 
necessarily forms as collapse end state and we need to find the 
conditions when the final singularity is 
necessarily covered within an event horizon (as hypothesized by 
the cosmic censorship), or when it will be naked with strong gravity 
regions being visible to faraway observers. In the case of a 
bounce or dispersal, no singularity of course need to form in the 
spacetime, a situation with which we do not concern 
ourselves here.

We can use the scaling freedom available for the radial 
co-ordinate $r$ to write $R=r$ at the initial epoch $t=t_i$. It 
is interesting
to note that, if we had wished to scale the radial co-ordinate at the
initial epoch as  $R(t_i,r)=r^{\beta}$, $\beta$ being any constant, 
then the {\it only} possible allowed value for $\beta$ is unity. 
Because in the case otherwise, either $R'$ would blow up at the center 
(which is not allowed as the Einstein's equations 
would require the metric functions to be at least $C^2$), or $R'$ would go
to zero at the center causing a shell-crossing singularity (which
we would like to avoid, by construction), which violates the 
regularity of the initial data.

We now introduce a function $v(t,r)$ as defined by,
\begin{equation}
v(t,r)\equiv R/r 
\label{eq:R}
\end{equation}
We then have $R(t,r)=rv(t,r)$, and 
\begin{eqnarray}
v(t_i,r)=1; & v(t_s(r),r)=0; & \dot{v}<0
\label{eq:v}
\end{eqnarray}
The time $t=t_s(r)$, that is $v=0$, corresponds to the shell-focusing 
singularity at $R=0$, which is a genuine spacetime singularity 
where all the matter shells collapse to a 
vanishing physical radius. The condition $\dot v< 0$ here 
corresponds to a continual collapse of the cloud. 
The description of shell-focusing singularity at $R=0$ in terms of 
the function $v(t,r)$ has several advantages. The physical radius 
goes to the zero value at the shell-focusing singularity, but we 
also have $R=0$ at the regular center of the cloud at $r=0$. This 
is to be distinguished from the genuine singularity at the collapse 
end state by the fact, for example, that the density and other 
physical quantities including the curvature scalars all remain 
finite at the regular center $r=0$, even though $R=0$ holds there. 
This is achieved, as we point out below, by a suitable behavior 
of the mass function, which should go to a vanishing value 
sufficiently fast in the limit of approach to
the regular center where (even though $R$ goes to zero)
the density must remain finite. On the other hand, when we use the
function $v(t,r)$, we note that
at $t=t_i$ we have $v=1$ on the entire initial surface, and then as the 
collapse evolves, the function $v$ continuously decreases to become 
zero only at the singularity $t_s(r)$, that is, $v=0$ uniquely 
corresponds to the genuine spacetime singularity at $R=0$.

>From the point of view of dynamic evolution of the initial data
prescribed at the initial epoch $t=t_i$, there are five arbitrary 
functions of the comoving shell-radius $r$ 
~\cite{gen4},
as given by,
\begin{eqnarray}
\nu(t_i,r)=\nu_0(r),&\psi(t_i,r)=\psi_0(r),& R(t_i,r)=r,\\\nonumber
\rho(t_i,r)=\rho_0(r),& p_r(t_i,r)=p_{r_0}(r),&\pt(t_i,r)=p_{\T_0}
\label{eq:initdata}
\end{eqnarray}
We note that not all the initial data above are mutually independent, 
because from equation (\ref{eq:ein2}) we get,
\begin{equation}
\nu_0(r)=\int_0^r\left(\frac{(N-2)(p_{\T_0}-p_{r_0})}
{r(\rho_0+p_{r_0})}-\frac{p_{r_0}'}{\rho_0+p_{r_0}}\right)dr
\label{eq:nu0}
\end{equation}
Thus, apart from the matter initial data describing the initial
density and pressure profiles, the rest of the initial data which is 
free is $\psi_0(r)$, which essentially describes the velocities of
the collapsing matter shells as we shall discuss later.

To ensure regularity of the initial data, the initial pressures 
must be taken to have physically reasonable behavior at the center. 
Considering that the total force at the center of the collapsing 
cloud should be zero, 
we have the gradients of initial pressures vanishing at the center.
Also, regularity of the initial data requires that we must have,
$\prz(0)-\ptz(0)=0$, that is, at the center the difference between the
radial and tangential pressure vanishes. The metric functions have
to be $C^2$ differentiable everywhere as per the requirements of the 
Einstein equations, and as seen from the equation for $\nu'$, the above
condition is implied by the requirement that $\nu'$ does not blow 
at the regular center.  This means that the matter 
should behave like a perfect fluid at the center of the cloud with 
the net force vanishing there. We note that these regularity conditions 
do not exclude the purely tangential pressure ($p_r=0$) or purely 
radial pressure collapse models ($p_\theta=0$) which we shall refer 
to later, because in those cases the above implies that 
$p_\theta \to 0$ or $p_r \to 0$ respectively, close to the center, 
where the matter then closely approximates dust. We note that these 
regularity conditions give us a sufficient condition for the 
regularity of the metric function $\nu_0(r)$ at any non-singular 
initial epoch. It follows from \e(\ref{eq:nu0}) that at the center 
of the cloud both $\nu_0$ and $\nu'_0$ go to zero. Hence 
$\nu_0(r)$ has the form,
\begin{equation}
\nu_0(r)=r^2g(r)
\label{eq:nu0form}
\end{equation} 
where, $g(r)$ is an arbitrary function which is at least $C^1$  
for $r=0$, and it is at least a $C^2$ function for $r>0$, as 
Einstein equations demand the metric functions to be at least 
$C^2$ everywhere. Another regularity condition frequently
used in collapse considerations is there are no trapped surfaces
at the initial surface from which the collapse begins.

We thus see that there are five total field equations with 
seven unknowns, $\rho$, $p_r$, $\pt$, $\psi$, $\nu$, $R$, and $F$, 
giving us the freedom of choice of two free functions. 
Selection of these free functions, subject to the weak energy 
condition and the given regular initial data for collapse at the 
initial surface, determines the matter 
distribution and metric of the space-time, and thus leads to a particular 
time evolution of the initial matter and velocity distributions. 
As we shall show, it turns out that given the matter initial profiles 
in terms of $\rho_0, p_{r_0}$ and $p_{\T_0}$, there exist rest of the 
initial data at $t=t_i$, and classes of solutions, which we find by 
means of explicit construction, which give either a black hole or a 
naked singularity as the end state of collapse. The outcome depends
on the nature of rest of the initial functions, and the classes 
of dynamical evolutions as allowed by the Einstein equations.

An important point to be noted here is, in the description above
we have made no mention so far on the equation of state that the matter 
must obey. Typically, these are of the form, $p_r=p_r(\rho)$ and 
$\pt=\pt(\rho)$. If these are specified, then there is no freedom 
left, and we have seven equations for seven variables. If we are to 
incorporate this right away, the only way to proceed to find the 
collapse end state would be to {\it assume} a specific 
equation of state that the matter must satisfy, and then to examine 
the collapse problem and the nature of the final singularity as 
resulting from the dynamical evolution as governed by the Einstein 
equations. There have been many collapse studies in past using such
an approach, e.g. for dust equation of state, perfect fluids etc.
The limitation of such an approach, however, has been that there is 
very little existing knowledge on what a realistic equation of state 
should be that the matter has to satisfy at the extreme high 
densities that a continual collapse  realizes in its advanced stages. 
For example, even for neutron star densities which are relatively 
low as compared to those of continual collapse, there is a great 
deal of uncertainty on the equation of state for such neutron 
matter. As a result, the neutron star mass limits are uncertain to 
that extent. Thus, specific or special assumptions used on the 
equation of state may turn 
out to be physically unrealistic or restrictive and untenable 
in the final stages 
of collapse. In fact, diametrically opposite views exist on the 
possible equation of state in very late stages of collapse. 
For example, while there are many arguments suggesting that pressures 
must play important role in the later stages of collapse, the 
opposite view is that in such late stages the matter must 
necessarily be dustlike (see e.g. 
~\cite{haga1}, \cite{pen1}) .

Under the situation, the path we take here is, we do {\it not} 
assume any specific or particular equation of state presently, and 
carry out our further considerations in a general way in terms of 
the allowed initial matter profiles, and the allowed dynamical 
evolutions of the Einstein equations, to determine the 
black hole and naked singularity end states for collapse. We then 
discuss subsequently, in Section V, the role that the equation of 
state will play towards further fine tuning the BH/NS outcomes 
as collapse end states. The advantage such an approach 
has is, first we write all the collapse equations in generality, 
and then only different subcases can be examined depending on 
the corresponding equation of state under consideration. As we 
shall show, various important subcases such as dust, perfect fluids 
and others can be included as special cases of the treatment 
given here.

In this paper our basic strategy is as described below.
We actually do not choose any explicit form of F(t,r), it is allowed 
to be completely general class, subject to regularity conditions.  
Till Section V, we do not make any special choices of the two free 
functions, that we refer to, but deal with and construct  
certain general classes, as determined by the differentiability 
conditions on the concerned functions (e.g. as specified by the equation 
(45) onwards) 
so that basically the singularity curve (which is defined as the time taken 
for a shell labelled `$r$' to reach the singularuty) is expandable upto 
at least first order, i.e. the singularity curve is regular enough so as to 
have a well-defined tangent, for these classes of evolutions. Then we show 
that these classes contains both BH and NS final states, as decided by 
the sign of the tangent of singularity curve.

In other words, given the matter initial data, we construct rest of the
functions so that in the continual collapse $(\dot R < 0)$, the final
singularity curve is regular (expandable) as above with a well-defined 
tangent. With that we then show, when the final singularity is visible, 
or covered in a black hole.
The key point is, once we have the matter initial data, we show the 
{\it existence} of classes of rest of the initial data and evolutions which 
are solutions to the Einstein equations, by explicit construction as 
specified here, so that the final state is either a BH or NS depending on 
the choice made of the rest of the initial data and evolutions. This is 
subject to energy conditions and other regularity conditions. We are of 
course not dealing with {\it all} 
possible classes of evolutions from the given matter initial data, which 
is not our purpsoe, but we just show the existence of the classes that 
lead the collapse to above final states.

For a given matter initial data, once we have chosen the rest of the 
initial data, and the evolutions as solutions to the Einstein equations, 
that clearly fixes the equation of state for the matter. That can be quite 
`exotic' at times, depending on the choices made. However, as shown explicitly 
in Section V, the construction given here does include well-known equations 
of state, such as dust, perfect fluids etc.

\section{Gravitational collapse of matter clouds}

The method we follow is outlined as below.
In the case of a black hole developing as collapse end state,
the spacetime singularity is necessarily hidden behind the event
horizon of gravity, whereas in the case of a naked singularity
developing there are families of future directed non-spacelike 
trajectories which terminate in the past at the singularity, 
which can in principle communicate information to faraway observers 
in the spacetime. The existence of such families confirms the NS formation, 
as opposed to a BH collapse end state. We study the
singularity curve produced by the collapsing matter, and it is 
shown that the tangent to the same at the central singularity at 
$r=0$ is related to the radially outgoing null geodesics from the 
singularity, if there are any. By determining the nature of the singularity
curve and its relation to the initial data and the classes of 
collapse evolutions, we are able to deduce whether the trapped surface 
formation in collapse takes place before or after the singularity. It is 
this causal structure of the trapped region forming during collapse
that determines the possible 
emergence or otherwise of non-spacelike curves from the singularity. 
This settles the final outcome of collapse in terms of either 
a BH or NS.

Given the matter initial profiles in terms of the functions
$\rho_0(r), p_{r_o}(r), p_{\T_0}(r)$ at the initial epoch $t=t_i$
from which the collapse commences, our purpose now is to construct and 
examine possible evolutions (classes of solutions to Einstein equations)
of such a matter cloud to investigate its final states.

While constructing the classes of solutions which give the 
collapse evolutions, given the matter initial data at $t=t_i$, we 
preserve as much generality as possible. Hence we allow the mass 
function $F(t,r)$ for the collapsing cloud to have a general form 
as given by,
\begin{equation}
F(t,r)=r^{N-1}\M (r,v)
\label{eq:mass}
\end{equation}
where $\M>0$ is at least a $C^1$ function of `$r$' for $r=0$, 
and at least a $C^2$ function for $r>0$.

It is to be noted that $F$ must have this general form 
follows from the regularity 
and finiteness of the density profile at the initial epoch $t=t_i$, 
and at all other 
later regular epochs before the cloud collapses to the final
singularity at $R=0$.
This requires, from the Einstein equation (17), that $F$ must behave 
as $r^{\left(N-1\right)}$ close to the 
regular center. Hence we note that since $\M$ is a general (at least 
$C^2$) function, the equation \e(\ref{eq:mass}) is not really any ansatz 
or a special choice, but quite a generic class of the mass profiles
for the collapsing cloud, consistent with and allowed by the regularity 
conditions. We thus make no special choice of $F$ but allow it
to be a general function as given by the above equation.
Then equation (\ref{eq:ein1}) gives, 
\begin{equation}
\rho(r,v)=\frac{N-2}{2}\left[\frac{(N-1)\M+
r\left[\M_{,r}+\M_{,v}v'\right]}{v^{N-2}(v+rv')}\right]
\label{eq:rho}
\end{equation}
and
\begin{equation}
\pr(r,v)=-\frac{(N-2)}{2}\left[\frac{\M_{,v}}{v^{N-2}}\right]
\label{eq:pr} 
\end{equation}
The regular density distribution at the initial epoch is given by,
\begin{equation}
\rho_0(r)=\frac{N-2}{2}\left[(N-1)\M(r,1)+r\M(r,1)_{,r}\right]
\label{eq:rho0}
\end{equation}
It is evident that, in general, as $v\rightarrow 0$, 
$\rho\rightarrow\infty$ and $\pr\rightarrow\infty$. That is, both the 
density and radial pressure blow up at the shell-focusing singularity.

It is seen that given any regular initial density
and pressure profiles for the matter cloud from which the collapse 
develops, there always exist energy profiles or velocity functions 
for the collapsing matter 
shells, and classes of dynamical evolutions as determined by the 
Einstein equations, so that the collapse end state would
be either a naked singularity or a black hole, depending  
on nature of the allowed choice. Thus, given the matter initial 
data at the initial surface $t=t_i$, these evolutions take the collapse 
to end either as a BH or NS depending on the choice 
of the class, subject to regularity and energy conditions.

To see this, we need to construct classes of solutions 
to Einstein equations to this effect. Let us define a suitably 
differentiable function $A(r,v)$ as follows,
\begin{equation}
\nu'(r,v)=A(r,v)_{,v}R'
\label{eq:A}
\end{equation}
That is, $A(r,v)_{,v}\equiv \nu'/R'$, and since at $t=t_i$
we have $R=r$ which gives 
$\left[A(r,v)_{,v}\right]_{v=1}=\nu_0'(r)$. Our main interest
here is in studying the shell-focusing singularity at $R=0$ which is 
the physical singularity where all the matter shells collapse to 
zero radius. Therefore we assume that there are no shell-crossing 
singularities in the spacetime where $R'=0$, and that the function 
$A(r,v)$ is well-defined. From equation (\ref{eq:nu0form}), we 
generalize and choose the form of $\nu(t,r)$ as the class 
given by,
\begin{equation}
\nu(t,r)=r^2g_1(r,v)
\label{eq:A2}
\end{equation}
where $g_1(r,v)$ is a suitably differentiable function and
$g_1(r,1)=g(r)$. It then follows that  $A(r,v)$ has the form,
\begin{equation}
A(r,v)=rg_2(r,v)
\label{eq:A4}
\end{equation}

From the regularity conditions and that the total force at the center of 
the collapsing cloud should be zero at any non-singular epoch, it is evident 
that both $g_1(r,v)$ and $g_2(r,v)$ shoud be well defined at $r=0$ and $v\ne 0$. 
In fact in some of the models discussed in section V, these functions are 
well behaved even at the singularity.
In general, it is possible that these functions may in fact blow up at the 
singularity in some cases. But, as we prove later, even if these functions 
do blow up, the singularity curve can still be well defined and expandable.

Some comments are in order on our assumption that $R'>0$,
that is, we have considered here the situation with no shell-crossing 
singularities. This is because, it is generally believed (see e.g.
\cite{shell1}~\cite{shell2})
that such singularities can be possibly removed from the spacetime 
as they are typically gravitationally weak, and because 
spacetime extensions have been constructed through the 
same in certain cases
\cite{clarke}. 
In contrast, in several physically reasonable collapse
models including dust and perfect fluids $R=0$ turns out to be a 
gravitationally strong curvature singularity.
Under the situation, we are interested
only in examining the nature of the shell-focusing singularities
at $R=0$, which are genuine curvature singularities arising 
at the termination of collapse, where the
physical radii for all collapsing shells vanish, and the
spacetime necessarily terminates without extension.

Specifically, $R'>0$ implies that we must have $v + rv'> 0$. 
Since $v$ is necessarily positive through out the collapse, it follow 
that this will be satisfied always whenever $v'$ is greater or equal to 
zero. Even when it is negative, the condition that the magnitude of 
$rv'$ should be less then that of $v$ is sufficient to ensure that 
there will be no shell-crosses. Later in this section we derive 
an expression for the quantity $v'$, in terms of the initial data and 
the other free evolutions as allowed by the Einstein equations. 
Hence it follows that we can specifically state the condition for 
avoidance of shell-crossings in terms of the behavior of these functions. 

%In particular, it turns out that whenever the singularity curve $t_s(r)$ 
%(which corresponds to $R=0$) is increasing at the center (or when it 
%decreases at a sufficiently slow rate) with a slope greater or equal to 
%zero at the origin, the shell-crossing singularities are avoided, 
%at least in the vicinity of the regular center $r=0$. We then have a 
%ball of finite radius around the regular center which contains 
%no shell-crossings till the final singularity formation 
%at $R=0$. 

Coming to the dynamical collapse evolutions,
using \e (\ref{eq:A}) in \e (\ref{eq:ein3}) we get,
as a class of solutions of Einstein's equations,
\begin{equation}
G(r,v)=b(r)e^{2rA(r,v)}
\label{eq:G}
\end{equation}
Here $b(r)$ is another arbitrary function of the shell radius $r$. 
By the regularity condition on the function $\dot{v}$ at the
center of the cloud we get the form of $b(r)$ as,
\begin{equation}
b(r)=1+r^2b_0(r)
\label{eq:veldist}
\end{equation}
where $b_0(r)$ is the energy distribution function for the 
shells. Using \e(\ref{eq:A}) in \e(\ref{eq:ein2}), we get,
\begin{equation}
(N-2)\pt=RA_{,v}(\rho+\pr)+(N-2)\pr+\frac{Rp_r'}{R'}
\label{eq:ptheta}
\end{equation}
In general, both the density and radial pressure blow up at 
the singularity, so the above equation implies that the tangential 
pressure would also typically blow up at the singularity.
Now using equations (\ref{eq:mass}), (\ref{eq:A}) and (\ref{eq:G}) 
in \e(\ref{eq:ein4}), we get,
\begin{equation}
R^{\frac{N-3}{2}}\dot{R}=-e^{\nu(r,v)}\sqrt{(1+r^2b_0)R^{(N-3)}e^{2rA(r,v)}
-R^{(N-3)}+r^{(N-1)}\M}
\label{eq:collapse}
\end{equation}
The negative sign in the RHS of the above equations corresponds 
to a collapse scenario where we have $\dot{R}<0$.
Defining a function $h(r,v)$ as,
\begin{equation}
h(r,v)=\frac{e^{2rA(r,v)}-1}{r^2}=2g_2(r,v)+{\cal O}(r^2)
\label{eq:h}
\end{equation}
the equation (\ref{eq:ein4}) becomes
\begin{equation}
v^{\frac{N-3}{2}}\dot{v}=-\sqrt{e^{(rA+\nu)}v^{(N-3)}b_0+e^{2\nu}
\left(v^{(N-3)}h+\M\right)}
\label{eq:dotv}
\end{equation}
Integrating the above equation with respect to $v$, we get,
\begin{equation}
t(v,r)=\int_v^1\frac{v^{\frac{N-3}{2}}dv}{\sqrt{e^{(rA+\nu)}v^{(N-3)}b_0+
e^{2\nu}\left(v^{(N-3)}h+\M\right)}}
\label{eq:scurve1}
\end{equation}
Note that the variable $r$ is treated as a constant in the above equation.
The above equation gives the time taken for a shell labeled $r$ to 
reach a particular {\it epoch} $v$ from the initial epoch $v=1$. 
Expanding $t(v,r)$ around the center of the cloud, we get,
\begin{equation} 
t(v,r)=t(v,0)+r\X(v)+{\cal O}(r^2)
\label{eq:scurve2}
\end{equation}
where the function $\X(v)$ is given as,
\begin{equation}
\X(v)=-\frac{1}{2}\int_v^1dv\frac{v^{\frac{N-3}{2}}(v^{(N-3)}b_1+h_1v^{(N-3)}
+\M_1(v))}{(v^{(N-3)}b_{00}+v^{(N-3)}h_0+\M_0(v))^{\frac{3}{2}}}
\label{eq:tangent1}
\end{equation}
wherein we have defined,
\begin{eqnarray}
b_{00}=b_0(0);&\M_0(v)=\M(0,v);&h_0=h(0,v) \nonumber \\
b_1=b_0'(0);&\M_1(v)=\M_{,r}(0,v);&h_1=h_{,r}(0,v)
\label{eq:notation}
\end{eqnarray}
Hence we see that the time taken for a shell labeled $r$ to reach
the spacetime singularity at $R=0$ (which is the {\it singularity curve}) 
is given as,
\begin{equation}
t_s(r)=\int_0^1\frac{v^{\frac{N-3}{2}}dv}{\sqrt{e^{(rA+\nu)}v^{(N-3)}b_0+
e^{2\nu}\left(v^{(N-3)}h+\M\right)}}
\label{eq:scurve3}
\end{equation}

As we want to consider here continual collapse, we focus 
only on those classes of solutions where $t_s(r)$ is finite and 
sufficiently regular. This means that the cloud collapses in a finite 
amount of time. In the physical situation of a continual collapse 
of a massive matter cloud in a finite amount of time the function 
$t_s(r)$ has to be of course finite. 
As for regularity, to check the existence conditions for 
a well defined, continuous and $C^2$ singularity curve, let us define 
a function $Q(r,v)$ as,
\begin{equation}
Q(r,v)=\frac{v^{\frac{N-3}{2}}}{\sqrt{e^{(rA+\nu)}v^{(N-3)}b_0+
e^{2\nu}\left(v^{(N-3)}h+\M\right)}}
\label{eq:q}
\end{equation}
Also we consider the following functions,
\begin{equation}
\phi_1(r)=\int_0^1Q(r,v)_{,r}dv\;;\;\phi_2(r)=\int_0^1Q(r,v)_{,rr}dv
\label{eq:phi}
\end{equation}
Let $\cal{A}$ be the rectangular area in $(r,v)$ plane defined by 
the lines,
\begin{equation}
r=0;r=\epsilon\;\;;v=0;v=1
\end{equation}
Now if the following conditions are satisfied 
~\cite{gibbons},
\begin{enumerate}
\item $Q(r,v)$ is a continuous function of $r$ and $v$ 
in $\cal{A}$,
\item $Q(r,v)_{,r}$ and $Q(r,v)_{,rr}$ are continuous functions 
of $r$ and $v$ in $\cal{A}$,
\item The integrals $\phi_1(r)$ and $\phi_2(r)$ converge uniformly 
in $\cal{A}$.
\end{enumerate}
then we can write,
\begin{equation}
\phi_1(r)=\frac{d}{dr}[t_s(r)]\;;\;\phi_2(r)=\frac{d^2}{dr^2}[t_s(r)]
\end{equation}
and this implies that the singularity curve $t_s(r)$ 
would be a well-defined $C^2$ function near the center.

We would like to emphasize here that regularity of different functions 
that makes $Q(r,v)$, namely $\nu$, $A$, and $M$, is sufficient but by 
no means necessary condition for the existance of a $C^2$ singularity 
curve. As we can easily see, even if these functions blow up at 
$r=0$, $v=0$, the function $Q$ may still be well defined. For example,
if we suppose these functions tend to infinity at the singularity 
the function $Q$ would tend to zero. For any regular collapse, the 
function $t_s(r)$ has to be finite and well-defined.

Several well-studied collapse models such as dust
collapse and others satisfy these or stronger conditions, and 
so the singularity curve is well-defined and expandable. 
Some examples of such singularity curves are as follows. In case of an 
N-dimensional dust collapse, the expression of the singularity curve 
would be
~\cite{gospsj1},
\begin{equation}
t_{s}(r)_{dust}=\int_0^1\frac{v^{\frac{N-3}{2}}dv}{\sqrt{\M(r)+v^{(N-3)}
b_{0}(r)}}
\label{eq:scurved}
\end{equation}
where $\M(r)$ and $b_{0}(r)$ are well defined $C^2$ functions of 
the comoving coordinate $r$ and well defined at $r=0$. It can then 
be easily seen that the singularity curve is differentiable at 
the center, with,
\begin{equation}
\frac{dt_s(r)}{dr}=-\frac{1}{2}\int_0^1\frac{v^{\frac{N-3}{2}}
(\M_1+v^{(N-3)}b_1)dv}
{(\M_0+v^{(N-3)}b_{00})^{\frac{3}{2}}}
\end{equation}
where we have defined,
\begin{eqnarray}
b_{00}=b_0(0),&\M_0=\M(0) \nonumber \\
b_1=b'(0),&\M_1=\M'(0)
\end{eqnarray}

Another such example is an N-dimensional 
{\it Einstein Cluster}, which describes a non-steady spherically 
symmetrical system of non-colliding particles moving in such a way 
that relative to a suitably moving frame of co-ordinates, 
their motion is purely transversal. In this case, the singularity 
curve is given by
~\cite{ashutosh},
\begin{equation}
t_s(r)_{Ec}=\int_{0}^{1}\frac{v^{\frac{N-3}{2}}\sqrt{v^2+
\frac{L(r)^2}{r^2}}\, dv}
{e^{\nu}\sqrt{b_{0}v^{N-1}-\left(\frac{L(r)^2}{r^4}\right)v^{N-3}+\M(r)
\left(v^2+\frac{L(r)^2}{r^2}\right)}}
\label{ec}
\end{equation}
Here $L(r)$ is a function of the radial coordinate $r$ only.
In the Newtonian limit, this function corresponds to the
angular momentum per unit mass of the system. Therefore
the function $L(r)$ is called as the
{\it specific angular momentum}, and has the form,
\begin{equation}
L(r)\equiv r^2l(r)
\end{equation}
Again, since $\M(r)$ and $b_{0}(r)$ and $L(r)$ are well-defined 
$C^2$ functions of the coordinate $r$, we see that the above singularity 
curve is well defined and differentiable at $r=0$

To generalize this and to give another explicit 
example, it can be shown that
given any initial data of the form (\ref{eq:initdata}), there always
exist classes of dynamical evolutions, which give rise to a well 
defined and 
differentiable singularity curve. Let us, by the freedom of choice 
of free functions, choose the evolution functions $\M(r,v)$ and 
$A(r,v)$ in the following way,
\begin{equation}
\M(r,v)=m(r)-p_r(r)v^{N-1},\; A(r,v)_{,v}=\nu_0(x)_{,x},\; x=rv
\end{equation}
From Einstein equations it is clear that for the above class
of evolutions the radial pressure remains {\it static}. However,
the tangential pressure blows up along with the density at the
singularity $v=0$ and is given by,
\begin{equation}
2p_{\theta}(r,v)= p_r+p_r'(R/R')+\nu_0(rv)_{,R}[\rho(r,v)+p_r]
\end{equation}
One can now easily check that the above class of evolutions 
admits a well-defined and differentiable singularity curve because 
both the functions $\M(r,v)$ and $A(r,v)$ are well-defined and 
$C^2$ at $r=0$ and $v=0$.

Once we have a singularity curve which is at least 
$C^2$, we can Taylor expand the function near the center as,
\begin{equation}
t_s(r)=t_{s_0}+r\X(0)+{\cal O}(r^2)
\label{eq:scurve4}
\end{equation}
where $t_{s_0}$ is the time when the central singularity 
at $R=0, r=0$ develops,
and is given as,
\begin{equation}
t_{s_0}=\int_0^1\frac{v^{\frac{N-3}{2}}dv}{\sqrt{v^{(N-3)}b_{00}+
v^{(N-3)}h_0+\M_0(v)}}
\label{eq:scurve5}
\end{equation}
From the above equation it is clear that for $t_{s_0}$ to be defined,
\begin{equation}
v^{(N-3)}b_{00}+v^{(N-3)}h_0+\M_0(v)>0
\label{eq:constraint}
\end{equation}
In other words, a continual collapse in finite time 
ensures that the above condition holds.
Also, from \e(\ref{eq:dotv}) and (\ref{eq:scurve2}) we get for 
small values of $r$, along constant $v$ surfaces,
\begin{equation}
v^{\frac{N-3}{2}}v'=\sqrt{(v^{(N-3)}b_{00}+v^{(N-3)}h_0+\M_0(v))}\X(v)
+{\cal O}(r)
\label{eq:vdash}
\end{equation}
It is now clear that the value of $\X(0)$ depends 
on the functions $b_0,\M$ and $h$, which in turn depend on 
the initial data at $t=t_i$, the dynamical variable $v$,
and the evolution function $A(r,v)$. 
Thus, a given set of initial matter distributions
and the dynamical profiles including the energy distribution 
of shells completely determine the tangent at the center
to the singularity curve.

\section{Apparent horizon and the nature of the singularity}

It is now possible to examine, given the matter initial data 
at the initial surface $t=t_i$, how the final fate of collapse 
is determined in terms of either a black hole or a naked singularity. 
If there are families of future directed non-spacelike trajectories 
reaching faraway observers in spacetime, which terminate in the past
at the singularity, then we have a naked singularity forming as the 
collapse final state and in the case otherwise when no such families 
exist and event horizon forms sufficiently earlier than the 
singularity to cover it, we have a black hole. This is decided by 
the causal behavior of the trapped surfaces developing in the 
spacetime during the collapse evolution, and the apparent horizon, 
which is the boundary of trapped surface region
in the spacetime.

In general, the equation of apparent horizon in a spherically 
symmetric spacetime is given as, 
\begin{equation}
g^{ik}\,R_{\, ,i}\,R_{\, ,k}=0
\label{eq:hor}
\end{equation}
Thus we see that at the boundary of the trapped region 
the vector $R_{\, ,i}$ is null. Substituting (\ref{eq:metric}) 
in (\ref{eq:hor}) we get,
\begin{equation}
R'^{2}\,e^{-2\psi} - \dot{R}^{2}\,e^{-2\nu}\,=\,0
\label{eq:hor1}
\end{equation}
Using equation (\ref{eq:ein4}) we can now write the 
equation of apparent horizon as,
\begin{equation}
\frac{F}{R^{N-3}}=1
\label{eq:apphorizon} 
\end{equation}
which gives the boundary of the trapped surface region of the 
spacetime. If the neighborhood 
of the center gets trapped prior to the epoch of singularity, then it 
is covered and a black hole results, otherwise it could be naked when  
non-spacelike future directed trajectories escape from it.

Thus the important point is to determine if there are any 
future-directed non-spacelike paths emerging from the singularity. 
To investigate this, and to examine the nature of the 
central singularity at $R=0, r=0$, let us consider the equation 
for outgoing radial null geodesics which is given by,
\begin{equation} 
\frac{dt}{dr}=e^{\psi-\nu}
\label{eq:null1}
\end{equation}
We want to examine if there would be any families of 
future directed null geodesics coming out of the singularity, 
thus causing a naked singularity phase as collapse endstate.
The singularity occurs at $v(t_s(r),r)=0$, i.e. $R(t_s(r),r)=0$. 
Therefore, if there are any future directed null geodesics 
terminating in the past at the singularity, we must have 
$R\rightarrow0$ as $t\rightarrow t_s$ along these curves. Now writing 
\e (\ref{eq:null1}) in terms of variables $(u=r^\alpha,R)$ , we have,
\begin{equation}
\frac{dR}{du}=\frac{1}{\alpha}r^{-(\alpha-1)}R'\left
[1+\frac{\dot{R}}{R'}e^{\psi-\nu}\right]
\label{eq:null2}
\end{equation}
In order to get the expression of the tangent to null geodesics 
emerging in the $(R,u)$ plane,
we choose a particular value of $\alpha$ such that the geodesic equation 
is expressed only in terms of known limits. 
For example if $\X(0)\ne0$, and the functions $\M$ and $h$ are well 
defined for $0\le r\le r_b$ and $0\le v\le 1$ we choose 
$\alpha=\frac{N+1}{N-1}$. 
Using \e (\ref{eq:ein4}) and considering $\dot{R}<0$, we then get
the null geodesic equation in the form,
\begin{equation}
\frac{dR}{du}=\frac{N-1}{N+1}\left(\frac{R}{u}+
\frac{v'v^{\frac{N-3}{2}}}{(\frac{R}{u})^{\frac{N-3}{2}}}\right)
\left(\frac{1-\frac{F}{R^{N-3}}}{\sqrt{G}[\sqrt{G}+\sqrt{H}]}\right)
\label{eq:null3}
\end{equation}

If the null geodesics do terminate at the singularity in 
the past with a definite tangent, then at the singularity the 
tangent to the geodesics have $\frac{dR}{du}>0$ in the $(u,R)$ plane, 
and must have a finite value. 
In the case of a massive singularity in dimensions greater than 
or equal to four,({\it i.e.} $F(t_s(r),r)>0$ for
$r\ne0$), all singularities for $r>0$ are 
covered since $\frac{F}{R^{N-3}}\rightarrow\infty$ and hence 
$\frac{dR}{du}\rightarrow-\infty$. This is when both the pressures
$p_r$ and $p_\theta$ are positive with the energy condition being
satisfied. Therefore in such a case only the central 
singularity at $R=0,r=0$ could be naked.

Hence we need to examine the central singularity at $r=0,R=0$
to determine if it is visible or not and to determine if 
there are any solutions existing to the outgoing null geodesics equation, 
which terminate in the past at the singularity, going to faraway
observers in future. The conditions are to be determined under which
this can happen. We also
note that, since the singularity curve and the evolution functions 
are regular, we can calculate the limit of the functions $H$, $G$ and 
$F/R$ at $r\rightarrow 0, t\rightarrow t_{s_0}$. From equation 
(\ref{eq:G}), as $A(r,v)$ is a well defined function, we can 
see that $G(t_{s_0},0)=1$. Also from equation (\ref{eq:dotv}) we 
see that at this point $H\approx r^2/v^{N-3}$. Calculating this 
limit on $t=t_{s_0}$ plane from equation (\ref{eq:vdash}), at 
the point $(t_{s_0},0)$, we have $H=0$. Hence we see from equation 
(\ref{eq:ein4}) that $F/R^{N-3}=0$ in this limit.

Let now $x_{0}$ be the tangent to the outgoing null geodesics 
in $(R,u)$ plane, at the central singularity, then it is given by,
\begin{equation}
x_0=\lim_{t\rightarrow t_s}\lim_{r\rightarrow 0} 
\frac{R}{u}=\left.\frac{dR}{du}\right|_{t\rightarrow t_s;r\rightarrow 0}
\end{equation}
To find out whether the null geodesic equation admits any 
solution of $x_{0}$ which is positive and finite at the central 
singularity, we can use the values of $H$, $G$ and $F/R$ 
at $(t_{s_0},0)$ in equation (\ref{eq:null3}). Also we use 
equation (\ref{eq:vdash}) to get the value of $v'v^{\frac{N-3}{2}}$, 
on $v=0$ surface at $r=0$ (that is, on the point $(t_{s_0},0)$). 
Thus solving \e (\ref{eq:null3}), we get,
\begin{equation}
x_0^{\frac{N-1}{2}}=\frac{N-1}{2}\sqrt{\M_0(0)}\X(0)
\label{eq:ruslope}
\end{equation}
and the equation of radial null geodesic emerging from the 
singularity is given by $R=x_0u$ in the $(R,u)$ plane, or in $(t,r)$ 
coordinates it is given by 
\begin{equation}
t-t_s(0)=x_0r^{\frac{N+1}{N-1}}
\end{equation}

It follows now that if $\X(0)>0$, then $x_0>0$, and we 
get radially outgoing null geodesics coming out from the singularity, 
giving rise to a naked central singularity.
However, if $\X(0)<0$ we have a black hole solution, as there will
be no such trajectories coming out. If $\X(0)=0$ then we will have 
to take into account the next higher order non-zero term in the 
singularity curve equation, and a similar analysis has to be 
carried out by choosing a different value of $\alpha$.

To show that the above is a necessary as well as sufficient 
condition for an outgoing radial null geodesic emerging form the 
singularity to exist, let us assume that such geodesics do exist 
and in the $(R,u)$ plane, it's equation is $R=x_0u$, and $x_0>0$. 
Then at the central singularity $(R=0,u=0)$, the tangent to such geodesic 
must be $x_0$. Also this tangent must be the root of the equation, 
\begin{equation}
\frac{dR}{du}-\frac{N-1}{N+1}\left(\frac{R}{u}+
\frac{v'v^{\frac{N-3}{2}}}{(\frac{R}{u})^{\frac{N-3}{2}}}\right)
\left(\frac{1-\frac{F}{R^{N-3}}}{\sqrt{G}[\sqrt{G}+\sqrt{H}]}\right)=0
\label{eq:null4}
\end{equation}
at the point $(R=0,u=0)$. This is possible if and only if 
\begin{equation}
x_0=\left[\frac{N-1}{2}\sqrt{\M_0(0)}\X(0)\right]
^{\frac{N-1}{2}}
\label{eq:ruslope1}
\end{equation}
and for the slope to be defined and positive we must have $\X(0)\ge0$.

Now to see that $\X(0)>0$ is a sufficient condition for 
the existence of an outgoing radial null geodesic emerging from the 
singularity, let us consider the case that the singularity curve has 
a positive 
tangent at the central singularity. Consider now the curve,
\begin{equation}
t-t_s(0)=\left[\frac{N-1}{2}\sqrt{\M_0(0)}\X(0)\right]^{\frac{N-1}{2}}
r^{\frac{N+1}{N-1}}
\label{eq:curve}
\end{equation}
Along this curve $t\rightarrow t_{s_0}$ as $r\rightarrow 0$. And as we 
have $\X(0)>0$, this curve is {\it outgoing} in the sense that 
$t$ increases as we increase $r$ along the curve. Let us now calculate the 
quantity $(-g_{tt}dt^2+g_{rr}dr^2)$ along this curve in the vicinity 
of the central singularity. Using (\ref{eq:vdash}), (\ref{eq:dotv}) and 
(\ref{eq:G}), we have for this curve at the point $( t_{s_0},0)$,
\begin{equation}
-e^{2\nu}dt^2+e^{-2rA}R'^2dr^2=
\frac{N+1}{N-1}\left[\frac{N-1}{2}\sqrt{\M_0(0)}\X(0)\right]^{N-1}
r^{\frac{2}{N-1}}(-dr^2+dr^2)=0
\end{equation}
That is, in the vicinity of the central singularity the curve 
(\ref{eq:curve}) is null. Thus we see that given any positive value 
of the tangent to the singularity curve at the central singularity, 
we can always find a {\it null} and {\it outgoing} curve terminating in 
the past at the central singularity, making the singularity naked.

We make below some remarks on the nature of the apparent horizon
and its relation with the visibility or otherwise of the singularity.
To find the equation of apparent horizon near the central 
singularity, let the time corresponding to a shell labeled by $r$ 
entering the apparent horizon, in term of the variable 
$v$, be $v_{ah}(r)$. Then from equation (\ref{eq:apphorizon}), 
we can easily see that $v_{ah}(r)$ is the root of the equation,
\begin{equation}
r^2\M(r,v)-v^{(N-3)}=0
\end{equation}
Now using \e(\ref{eq:scurve1}), we get the the equation 
for apparent horizon in $(t,r)$ plane as,
\begin{equation}
t_{ah}(r)=t_s(r)-\int_0^{v_{ah}(r)}\frac{v^{\frac{N-3}{2}}dv}
{\sqrt{{e^{(rA+\nu)}v^{(N-3)}b_0+
e^{2\nu}\left(v^{(N-3)}h+\M\right)}}}
\label{eq:apphorizon1}
\end{equation}
It is obvious that the necessary condition for the 
existence of a locally naked singularity is that the apparent horizon 
curve must be an {\it increasing} function at the central singularity, 
in the lowest power of $r$.

We note that in the above the functions $h$ and ${\cal M}$ 
are expanded with respect to $r$ around $r=0$ and the first-order 
terms are considered. At times, however, these are assumed to be
expandable with respect to $r^{2}$, and it is 
argued that such smooth functions would be physically more relevant.
Such an assumption comes from the analyticity with respect to the  
local Minkowskian coordinates 
(see e.g. \cite{gospsj2}), 
and it is really
the freedom of definition mathematically. We may remark that the 
formalism as discussed above would work for such smooth functions 
also, which is a special case of the above discussion.

We thus see how the initial data 
in terms of the free functions available determine the 
BH/NS phases as collapse end states, because $\X(0)$ is determined
by these initial and dynamical profiles as given by 
\e(\ref{eq:tangent1}). It is clear, therefore,
that given any regular initial density and pressure profiles for
the matter cloud from which the collapse develops, we can always 
choose velocity profiles so that the end state of the collapse would
be either a naked singularity or a black hole, and vice-versa.

We believe numerical work on collapse models may provide further 
insights into these interesting dynamical phenomena, and especially 
when collapse is non-spherical, which 
remains a major open problem to be considered
~\cite{nonsph1}-\cite{nonsph13}. 
Numerical and some analytical works have been done in recent 
years on spherical scalar field collapse 
~\cite{scalar1}-\cite{scalar9}  
and also on some perfect fluid models 
~\cite{fluid1}-\cite{fluid9}. 
While we have worked out explicitly here the
emergence of null geodesics from the singularity, thus showing it to be
naked (or otherwise) in an analytic manner, the numerical
simulations generally discuss the formation or otherwise of
trapped surfaces and apparent horizon, and such considerations
may possibly break down 
closer to the epoch of actual singularity formation. In that case,
this may not allow for actual detection of BH/NS end 
states, whereas important insights on critical phenomena 
and dispersal have already been gained through numerical methods. 
Probably, a detailed numerical investigation of the structure of 
null geodesics in collapse models may provide further 
insights here.

\section{Equation of State and energy conditions}

As stated above, we work here with {\it type I} matter fields, 
which is a rather general form of matter. However, it is
important to note that suitable care must be taken in interpreting 
these results. While we have shown that the initial data and dynamical 
evolutions chosen do determine the BH/NS end states for collapse, the 
point is, actually, all these dynamical variables are not explicitly 
determined by the initial data given at the initial epoch (note 
that $v$ plays the role of a time coordinate here). Hence these functions 
are fully determined only as a result of time development of the 
system from the initial data provided we have the relation between 
the density and pressures, that is a given `equation of state'.

In principle, it is possible to choose these functions freely (e.g.
the matter and velocity profiles at the initial epoch and the 
dynamical evolutions such as $F(v,r)$ and $\nu(v,r)$), only 
subject to an energy condition and regularity, which then fully
determines the collapse evolution. One can then calculate the 
energy density, and the radial and tangential 
pressures for the matter. However, in that case, the resultant 
`equation of state' could be quite strange in general. If any equation 
of state of the form $p_r=f(\rho)$ and/or $p_\theta=g(\rho)$
is given, then it is clear from equation (\ref{eq:ein1}) and (\ref{eq:ein2}) 
that there would be a constraint on the otherwise arbitrary function 
$\M$ and $A$, specifying the required class, if the solution of the 
constraint equation exists. It is certainly true that, presently 
we have practically very little idea on what kind of an equation of
state should the matter follow, especially at very high densities
and closer to the collapse end states, where we are already dealing 
with ultra-high energies and pressures. Hence if we allow for the 
possibility that we could freely choose the property of the matter 
fields as above, or the equation of state, then our analysis is 
certainly valid and give several useful conclusions on possible
collapse end states. In such a case, it is also possible that the 
chosen equation of state will be in general such that the pressures 
may explicitly depend not only on the energy density, but also on 
the time coordinate.

All the same, it is important to point out that the analysis 
as given above in fact does include several well-known equations of
state and useful classes of collapse models, also satisfying 
the energy conditions throughout the collapse, as we  
demonstrate below.

\subsection{Dust collapse}

The idealized class of dust collapse models where the 
pressures are taken to be vanishing has been studied extensively
so far and has yielded many important insights on collapse
evolutions.
In this special case, the Einstein equations can 
be solved completely to get the N-dimensional generalization 
of the usual Tolman-Bondi-Lemaitre (TBL) dust collapse 
metric 
(\cite{TBL1}-\cite{TBL3}) 
and it is given as,
\begin{equation} 
ds^2=dt^2-\frac{R^{'2}}{1+r^2b_0(r)}dr^2-R^2(t,r)\dw
\label{eq:tbl1}
\end{equation}
The  equations of motion are given by,
\begin{equation}
\frac{(N-2)F'}{2R^{(N-3)}R'}=\rho
\label{eq:eqnofmotion1}
\end{equation}
and
\begin{equation}
\rdot^2=\frac{\F}{R^{(N-3)}}+\f
\label{eq:eqnofmotion}
\end{equation}
In the case of dust, the mass function must be $F=F(r)$, and hence 
regularity condition implies that,
\begin{equation}
\F=r^{(N-1)}\M(r)
\label{eq:forms1}
\end{equation}
The energy condition here gives  $0<r<r_b$, and so we must have 
$\M(r)\ge0$ and $3\M+r\M_{r}\ge0$.
In this case, as we have already seen in section III, 
the function $\X(v)$ is given as,
\begin{equation}
\X(v)=-\frac{1}{2}\int_v^1\frac{v^{\frac{N-3}{2}}(\M_1+v^{(N-3)}b_1)dv}
{(\M_0+v^{(N-3)}b_{00})^{\frac{3}{2}}}
\label{eq:tangentd}
\end{equation}
and the time taken for the central shell to reach the 
singularity is given by,
\begin{equation}
t_{s_0}=\int_0^1\frac{v^{\frac{N-3}{2}}dv}{\sqrt{\M_0+v^{(N-3)}b_{00}}}
\label{eq:scurve3d}
\end{equation}
It is now seen clearly that any given sets of density and velocity 
profiles at the initial epoch completely determine the tangent to the 
singularity curve at the central singularity. 
Also the equation (\ref{eq:ruslope}) becomes 
\begin{equation}
x_0^{\frac{N-1}{2}}=\frac{N-1}{2}\sqrt{\M_0}\X(0)
\end{equation}

It therefore follows that, given any specific density profile 
of the collapsing dust cloud, we can always choose a velocity profile 
so that the end state of the collapse would be either a naked 
singularity or a black hole depending on the choice made, such that 
energy conditions are satisfied throughout the collapse. The converse
also holds, namely one can choose a given velocity profile for
the cloud at the initial epoch, and then there are density profiles
which will lead the collapse to either of the BH/NS final states,
and these conclusions hold irrespective of the number of dimensions
of the spacetime. Hence we see that our results
unify and generalize the earlier results of dust collapse
\cite{dust1}-\cite{dust7}. 
Basically the point that follows here is that, given an initial
density profile for the collapsing cloud, the space of velocity
profile functions is divided into the regions that lead the
collapse either to a black hole or naked singularity evolution,
depending on the choice made, and the converse holds similarly.

\subsection{Collapse with static radial pressure}

While the dust equation of state discussed above is fairly
standard and extensively used, it is widely believed that pressures
could play an important role in gravitational collapse 
considerations. We discuss below a class of collapse models
with non-zero pressures, which is however idealized in the
sense that while the tangential pressure can be arbitrary,
the radial pressure is taken to be static.

As we already pointed out in section III, if we consider 
the classes of collapse in which the radial pressure remains static, 
the constraint equation for $\M$ has the following solution,
\begin{equation}
\M(r,v)=m(r)-p_r(r)v^{N-1}
\end{equation}
In addition to this if there is an equation of state of the form 
$p_{\theta}=f(\rho)$, then that gives the constraint equation 
for the function $A(r,v)$ in the following way.
\begin{equation}
2f(\rho)= p_r+p_r'(R/R')+A(r,v)_{,v}[\rho(r,v)+p_r]
\end{equation}
For this class of models the energy conditions are given by,
\begin{equation}
(N-1)[m-p_rv^{N-1}]+r[m_{,r}-p_{r_{,r}}v^{N-1}-(N-1)p_rv^{N-2}v']\ge 0
\end{equation}
\begin{equation} 
(N-1)[m-p_rv^{N-1}]+r[m_{,r}-p_{r_{,r}}v^{N-1}-(N-1)p_rv^{N-2}v']+p_rR^2R'\ge 0
\end{equation}
\begin{equation}
\frac{\rho}{2}+\frac{1}{2}\left[(\rho+p_r)(A_{,v}+1)+p_r'\frac{R}{R'}\right]\ge 0
\end{equation}
As shown in 
~\cite{gospsj3} 
there exist classes of functions $m$, $p_r$ and $A$ such that 
naked singularity is the end state for the collapse and also the 
above three energy conditions are satisfied.
It follows that, given initial matter profiles, there exist
classes of collapse evolutions satisfying the energy conditions
as we see above, such that either of the BH/NS endstates can 
result subject to above equation of state.

\subsection{Isentropic perfect fluid with a linear equation of state}

Perfect fluids have been widely used in astrophysical
considerations and a linear equation of state is well-studied. We
discuss below how the formalism outlined here apply to this case
to find the BH/NS configurations as a perfect fluid collapse
end states.

For an {\it isentropic} perfect fluid, whose pressure is a 
linear function of the density only, the equation of state of the 
collapsing matter is given by,
\begin{equation}
p_r(t,r)=p_\theta(t,r)=k\rho(t,r)
\label{eq:pf}
\end{equation}
where $k\in[-1,1]$ is a constant. The case $k=0$ gives the
dust case we discussed above and $k=1$ is the stiff fluid case. 
Let us at present consider only the case of positive pressures. 
In that case $k>0$ and the energy conditions give,
\begin{equation}
M_{,v}<0
\end{equation}
>From the above equation of state and the Einstein equations we can 
immediately see that the function $\M$ is now the solution of the 
equation
\begin{equation}
(N-1)k\M + kr\M_{,r} + Q(r,v)\M_{,v}=0  
\label{eq:eos}  
\end{equation}
where,
\begin{equation}
Q(r,v)=(k+1)rv'+v
\label{eq:q1}
\end{equation}  
Now the above equation (\ref{eq:eos})
has a general solution of the form 
\cite{diff},  
\begin{equation}
{\cal F}(X,Y)=0
\label{eq:solution}
\end{equation}  
where $X(r,v,\M)$ and $Y(r,v,\M)$ are the solutions
of the system of equations,
\begin{equation}
-\frac{d\M}{(N-1)k\M}=\frac{dr}{kr}= \frac{dv}{Q}
\label{eq:auxilliary}
\end{equation}
Thus we can easily see that equation (\ref{eq:eos}) admits 
classes of solutions when $v'>0$. Also solving the equation for 
the central shell $r=0$, with boundary conditions 
$\rho\rightarrow\infty$ as $v\rightarrow 0$, we get,
\begin{equation}
\M(0,v)=\frac{m_0}{v^{(N-1)k}}
\end{equation}
By choosing $m_0>0$, we can make the central shell to 
satisfy the energy condition $\rho(t,0)>0$ for all epochs. Then 
by the continuity of the density function, we can say that 
there exists an $\epsilon$-ball around the central shell 
for which $v'(t,r)>0$ and also $\rho(t,r)>0$. But as know, at 
the central singularity $\sqrt{v}{v'}\approx\X(0)$, hence 
this implies that we can have classes of solutions which satisfy 
the energy conditions and also admits a naked singularity 
as the collapse end state.

For further discussion on perfect fluid collapse, and the
details of black hole and naked singularity formation, we refer to 
~\cite{fluid1}-~\cite{fluid9},
and references there in.

It is seen from the above that several well-known classes
of collapse models form subcases of the consideration given
here. Along with these well-known models, the above analysis 
would work for any other models with other equations of state, if 
that permit solutions to the constraint equations on $\M$ and $A$. 
Hence it follows that the considerations above provide an 
interesting framework for the study of dynamical collapse, 
which is one of the most important open problems in gravity 
physics today.

\section{The case of 2+1 dimensional collapse}

Several studies on gravitational collapse scenarios 
in $(2+1)$ dimensional spacetimes, have been carried out by various 
authors
~\cite{211}-~\cite{214}.
These provide interesting toy models which may provide quite
important insights from the perspective of quantum gravity. 
This is because in 
$(2+1)$ dimensional spacetimes there is no gravity outside matter. 
Also the spacetime metric is always conformally flat as the 
{\it Weyl tensor} vanishes identically everywhere. The situation in
$(3+1)$ and higher dimensions is far more complicated as compared
to this.

To investigate the final outcome in $(2+1)$ dimensional 
collapse, let us consider the geometry of the trapped surfaces 
in this case in some detail. From equation (\ref{eq:apphorizon}) 
we see that the equation of apparent horizon in this case 
is given by,
\begin{equation} 
F(t,r)=1
\label{eq:apphorizon2}
\end{equation}     
It is interesting to note that the geometry of trapped 
surfaces here is completely determined by the mass function of 
the cloud, and is independent of the area radius of the collapsing 
shells. If the mass function of the collapsing configuration is 
bounded from above, say for example with $F(t,r)<1$ for 
$t\in[-\infty,t_s(r)]$, and $r\in[0,r_b]$ (where $r_b$ is the boundary 
of the collapsing cloud), we then see that the trapping does not occur 
and the complete singularity that forms as collapse end state is 
necessarily visible to an outside observer. This is 
strikingly different from four or higher dimensional cases 
where a massive singularity is always trapped.

An interesting subcase of this situation is that of $(2+1)$ 
dimensional dust collapse, when $F(t,r)=F(r)$ with the mass function
having no time dependence. Here we see that the initial mass of the 
collapsing cloud completely determines the final outcome in terms of 
BH or NS. Clouds with small enough mass always form a visible singularity, 
whereas for larger masses a trapped region is present at all epochs. 
However, as demanded by the regularity conditions, we must avoid
trapped surfaces on the initial surface $t=t_i$ where the collapse
commences. In that case, for the dust collapse case, there are no 
trapped surfaces developing at all at any other later epochs till 
the singularity formation, and as a result the $(2+1)$ dust collapse 
always necessarily produces a visible naked singularity, as 
opposed to BH/NS phases obtained in usual four dimensional dust 
collapse which we discussed earlier.

\section{Exterior spacetime and matching conditions}

To complete a collapse model, we need to match the
interior spacetime to a suitable exterior spacetime. As we are
interested here in modelling collapse of astrophysical objects (such as 
massive stars), we have assumed the matter to have compact support
at the initial surface, with the boundary of the cloud being 
at some $r=r_b$. If one assumes the pressures at the boundary of 
the cloud to be vanishing, then it is always possible to match the 
interior spacetime with an empty Schwarzschild exterior. However, 
in all cases it may not be possible to make the pressures at the 
boundary of the cloud to vanish. Hence we outline here the procedure 
to match the interior with a general class of exterior metrics, 
which are the generalized Vaidya spacetimes 
~\cite{match1} and \cite{match2}, 
at the boundary hypersurface $\Sigma$ given by $r=r_b$. 
For the required matching we use the {\it Israel-Darmois} conditions 
(\cite{match4} and \cite{match5}), 
where we match the first and second fundamental forms, 
(the metric coefficients
and the extrinsic curvature respectively) at the boundary
of the cloud.

Whereas the procedures used below are standard, we shall
describe the particular case treated here in some detail so as to give 
the exact picture of the overall collapse scenario emerging.  
We note a useful fact that since we are matching
the second fundamental form $K_{ij}$, there is {\it no} surface stress
energy or surface tension at the boundary (see e.g. 
~\cite{match6}).
The metric just inside $\Sigma$ is,
\begin{equation}
ds^2_{-}=-e^{2\n}dt^2+e^{2\s}dr^2+R^2(t,r)d\Omega^2
\label{eq:metricgen}
\end{equation}
which describes the geometry of the collapsing cloud. The
metric in the exterior of $\Sigma$ is given by,
\begin{equation}
ds^2_{+}=-\left(1-\frac{2M(r_v,V)}{r_v}\right)dV^2-2dVdr_v+r_v^2d\Omega^2
\label{eq:metricvaidya}
\end{equation} 
where $V$ is the retarded (exploding) null co-ordinate and $r_v$
is the Vaidya radius. Matching the area radius at the boundary we get,
\begin{equation}
R(r_b,t)=r_v(V)
\label{eq:radius}
\end{equation}  
Then on the hypersurface $\Sigma$, the interior and exterior 
metrics are given by,
\begin{equation}
ds^2_{\Sigma-}=-e^{2\nu(t,r_b)}dt^2+R^2(t,r_b)^2d\Omega^2
\label{eq:metric3}
\end{equation} 
and
\begin{equation}
ds^2_{\Sigma+}=-\left(1-\frac{2M(r_v,V)}{r_v}+2\frac{dr_v}{dV}\right)dV^2
+r_v^2d\Omega^2
\label{eq:metric4}
\end{equation}
Matching the first fundamental form gives,
\begin{eqnarray}
\left(\frac{dV}{dt}\right)_\Sigma=\frac{e^{\nu(t,r_b)}}
{\sqrt{1-\frac{2M(r_v,V)}{r_v}
+2\frac{dr_v}{dV}}};&\left(r_v\right)_\Sigma=R(t,r_b)
\label{eq:match1}
\end{eqnarray}

Next, to match the second fundamental forms (extrinsic curvatures)
for the interior and exterior metrics, we note that the normal to 
the hypersurface $\Sigma$, as calculated from the interior metric, 
is given as,
\begin{equation}
n^i_{-}=\left[0,e^{-\psi(r_b,t)},0,0\right]
\label{eq:n1}
\end{equation} 
and the non-vanishing components of the normal as derived from the 
generalized Vaidya spacetime are,
\begin{equation}
n^V_{+}=-\frac{1}{\sqrt{1-\frac{2M(r_v,V)}{r_v}+2\frac{dr_v}{dV}}}
\label{eq:n2}
\end{equation} 
\begin{equation}
n^{r_v}_{+}=\frac{1-\frac{2M(r_v,V)}{r_v}+\frac{dr_v}{dV}}
{\sqrt{1-\frac{2M(r_v,V)}{r_v}+2\frac{dr_v}{dV}}}
\label{eq:n3}
\end{equation} 
Here the extrinsic curvature is defined as,
\begin{equation}
K_{ab}=\frac{1}{2}{\cal L}_{\bf n}g_{ab} 
\label{eq:k1}
\end{equation} 
That is, the second fundamental form is the Lie derivative
of the metric with respect to the normal vector ${\bf n}$.
The above equation is equivalent to,
\begin{equation}
K_{ab}=\frac{1}{2}\left[g_{ab,c}n^c+g_{cb}n^c_{,a}
+g_{ac}n^c_{,b}\right]
\label{eq:k2}
\end{equation}\\ 
Now setting $\left[K_{\theta\theta}^{-}-K_{\theta\theta}^{+}
\right]_{\Sigma}=0$
on the hypersurface $\Sigma$ we get,
\begin{equation}
RR'e^{-\psi}=r_v\frac{1-\frac{2M(r_v,V)}{r_v}+\frac{dr_v}{dV}}
{\sqrt{1-\frac{2M(r_v,V)}{r_v}+2\frac{dr_v}{dV}}}
\label{eq:match2}
\end{equation} 
Simplifying the above equation using equation (\ref{eq:match1}) 
and the Einstein equations, we get,
\begin{equation}
F(t,r_b)=2M(r_v,V)
\label{eq:match3}
\end{equation} 
Using the above equation and (\ref{eq:match1}) we now get,
\begin{equation}
\left(\frac{dV}{dt}\right)_\Sigma=\frac{e^\nu(R'e^{-\psi}+\dot{R}e^{-\nu})}
{1-\frac{F(t,r_b)}
{R(t,r_b)}}
\label{eq:match4}
\end{equation}
Finally, setting $\left[K_{\tau\tau}^{-}-K_{\tau\tau}^{+}\right]_{\Sigma}=0$,
where $\tau$ is the proper time on $\Sigma$,  
we get,
\begin{equation}
M(r_v,V)_{,r_v}=\frac{F}{2R}+\frac{Re^{-\nu}}{\sqrt{G}}\sqrt{H}_{,t}
+Re^{2\nu}\nu'e^{-\psi}
\label{eq:match5}
\end{equation}
Any generalized Vaidya mass function $M(v,r_v)$, which satisfies
equation (\ref{eq:match5}) will then give a unique exterior spacetime
with required equations of motion given by other matching conditions,
(\ref{eq:match3}), (\ref{eq:match4}) and (\ref{eq:radius}).

To see that the set of all such functions $M(v,r_v)$ is non-empty,
we have the examples of a charged Vaidya spacetime $M=M(V)+Q(V)/r_v$, 
and the anisotropic 
de-Sitter spacetime $M=M(r_v)$ as two different solutions of the
equation (\ref{eq:match5}) (see for example 
~\cite{match2} and ~\cite{match3}). 
This gives two unique exterior spacetimes, both of which are subclasses
of the generalized Vaidya metric.

\section{Concluding remarks}

We make here some remarks towards a conclusion and note some
unresolved and open issues.

1. Towards investigating endstates of a continual gravitational 
collapse, given a general {\it type I} matter field and
given the matter initial profiles at the initial surface from which
the collapse develops, we constructed here classes of solutions to
the Einstein equations such that the collapse evolution goes to
the formation of either of a black hole or naked singularity endstate, 
depending on the choice of the evolution made and choice of rest of 
the initial data functions such as velocities of the collapsing shells. 
This is subject to satisfying energy condition,
and several reasonable and important equations of state are included 
in the framework here. It also becomes clear that in higher dimensions 
also both black holes and naked singularities can occur as collapse
endstates.

2. We like to note that what we have deduced here is the 
occurrence of a {\it locally naked singularity} only, as opposed 
to that of a {\it globally naked singularity}. That is, we show 
when the null geodesics escape from the spacetime singularity,
going out in the future, but we do not address the question of when 
they go out of the boundary of the matter cloud. It is possible, 
in principle, that the singularity is only locally naked and trajectories 
do come out but they all fall back into the singularity again, 
without going out of the boundary of the star, thus not being 
globally visible.  

This issue is still not studied for the general class of models 
such as the classes we considered here and may be of interest. 
However, for the case of dust models this has been studied in 
some detail and it is shown that whenever the singularity is locally naked, 
one can always choose the classes of the mass and energy functions 
suitably, as one moves away from the center, in such a manner 
that the singularity becomes globally visible. The point is, while 
the local visibility of the central singularity is basically 
decided by the conditions near the center, the global visibility 
really depends on the overall behavior of these functions within
the matter cloud, away from the center. This we are still free to 
choose. In other words, for the dust collapse models once the 
singularity is locally visible, there are always classes of functions 
which we can choose so as to make it globally visible.

Another important related point here is, as such there is no scale
in the problem, and the size of the collapsing cloud could be quite
large. In such a case, even if the singularity is only locally visible,
still it can be seen for a long enough time by the observers. 
Thus, in principle, a locally naked singularity is also as serious
violation of the cosmic censorship as a globally visible singularity,
and there may not be a qualitative difference in the two cases
in many situations of physical interest.

3. Another important and interesting issue frequently mentioned
regarding occurrence of naked singularities in gravitational collapse 
is their genericity and stability. It is argued that if these are
not generic or stable, these need not be taken seriously. This is 
interesting because in general relativity there is no well-defined notion 
or criteria available for stability or genericity, which one can then 
apply and test for a given model to ascertain these. On the other hand, 
a consideration of this issue is important all the same in that,
depending on the collapse situation under discussion one would like to 
formulate these notions in some way to examine if the naked singularities 
developing as collapse endstates are `generic' or `stable' in some 
suitable sense. This would typically involve taking into account the
topology and metric of the concerned function spaces which define the
given collapse scenario. Without discussing this here in further detail, 
we note that since we have formulated here the collapse in a general 
manner, for general physically reasonable matter fields, this
classes constructed here may provide a good arena to explore and test
these important issues for the cosmic censorship hypothesis.

4. A related issue of course is that of non-spherical collapse.
Our considerations here are restricted to spherical collapse, and 
the question remains open regarding the final end state of a non-spherical
collapse. So one can ask if the conclusions available in spherically
symmetric collapse remain the same and stable under possible non-spherical 
perturbations. 

Even though some non-spherical collapse models have been discussed
and investigated so far such as the Szekeres quasi-spherical collapse
or some cylindrical collapse models
~\cite{nonsph1}-~\cite{nonsph13},
these are somewhat restricted in nature. 
A strong theorem about the formation of trapped surfaces in 
cylindrically symmetric spacetimes is given in 
~\cite{nonsph12}.
Recently, a numerical construction of naked singular solutions with the 
cylinderical symmetry was done in  
~\cite{nonsph13}.
It is not clear as yet if this issue can be approached really 
in an analytic manner, and possibly detailed numerical simulations of 
the collapsing stars could be the answer.

5. A question that is frequently asked in connection to the
occurrence of naked singularities as collapse endstates is that
how to understand this phenomena physically. A naked singularity 
signifies the escape of light and particle trajectories from the 
ultra-dense spacetime regions. However, the gravity must become so 
strong in these regions. In such a case, how can any thing escape at all 
from such a region is the question. Thus, while a black hole which 
is a region from which not even light would escape, may appear to be 
the only physically reasonable outcome in such situations, formation of
a naked singularity in collapse may appear to be counter-intuitive.

The point that comes out from considerations such as ours is 
that the naked singularities are more an artifact of general relativity, 
rather than that of a purely Newtonian physics. Even though the matter
density grows higher and higher without bound and blows up closer to 
a spacetime singularity, which would denote the growth of attractive forces of
gravity, there are other important factors which are purely general
relativistic effects which can delay the formation of trapped 
surfaces governing the trapping of light.

An interesting effect that does this is the spacetime shear. It 
is intriguing to find that the physical agencies such as the spacetime 
shear, and  related inhomogeneities in matter density distribution 
within a dynamically collapsing cloud, could naturally delay the 
formation of trapped surfaces during gravitational collapse 
~\cite{shear1}, ~\cite{shear2}.
In other words, such physical factors do naturally give rise 
to naked singularity phases in the collapse where the
formation of apparent horizon and the trapped surfaces is delayed.
Even though the matter densities are arbitrarily large and growing, 
the shear could distort the trapped surface geometry in such a 
manner so as to avoid the trapping of light and facilitates the
escape of null rays from such ultra-dense regions.

\end{document}